\documentclass[twocolumn,dutch,british,english,aps,prb,groupedaddress]{revtex4-1}
\usepackage[T1]{fontenc}
\usepackage[latin9]{inputenc}
\usepackage{textcomp}
\usepackage{amsmath}
\usepackage{amssymb}
\usepackage{graphicx}
\usepackage{esint}

\makeatletter

\DeclareRobustCommand{\greektext}{%
  \fontencoding{LGR}\selectfont\def\encodingdefault{LGR}}
\DeclareRobustCommand{\textgreek}[1]{\leavevmode{\greektext #1}}
\DeclareFontEncoding{LGR}{}{}
\DeclareTextSymbol{\~}{LGR}{126}
\newcommand{\lyxmathsym}[1]{\ifmmode\begingroup\def\b@ld{bold}
  \text{\ifx\math@version\b@ld\bfseries\fi#1}\endgroup\else#1\fi}

\providecommand{\tabularnewline}{\\}

\@ifundefined{textcolor}{}
{%
 \definecolor{BLACK}{gray}{0}
 \definecolor{WHITE}{gray}{1}
 \definecolor{RED}{rgb}{1,0,0}
 \definecolor{GREEN}{rgb}{0,1,0}
 \definecolor{BLUE}{rgb}{0,0,1}
 \definecolor{CYAN}{cmyk}{1,0,0,0}
 \definecolor{MAGENTA}{cmyk}{0,1,0,0}
 \definecolor{YELLOW}{cmyk}{0,0,1,0}
 }

\usepackage{amsfonts}\usepackage{longtable}

\makeatother

\usepackage{babel}

\makeatother

\usepackage{babel}
\begin{document}

\title{Mechanisms of nonthermal destruction of the superconducting state
and melting of the charge-density-wave state by femtosecond laser
pulses}

\author{L. Stojchevska$^{1}$, P. Kusar$^{1}$, T. Mertelj$^{1}$, V. V.
Kabanov$^{1}$, Y. Toda$^{1,2}$, X. Yao$^{3}$, and D. Mihailovic$^{1}$}

\affiliation{$^{1}$Complex Matter Department, Jozef Stefan Institute, Jamova
39, 1000 Ljubljana, Slovenia}

\affiliation{$^{2}$Department of Physics, Hokkaido University, Sapporo, Japan}

\affiliation{$^{3}$Department of Physics, Shanghai Jiao Tong University, Shanghai
200030, China}

\date{\today}
\begin{abstract}
The processes leading to nonthermal condensate \foreignlanguage{british}{vaporization}
and charge-density wave (CDW) melting with femtosecond laser pulses
is systematically investigated in different materials. We find that
\foreignlanguage{british}{vaporization} is relatively slow ($\tau_{v}\sim1$
\foreignlanguage{dutch}{ps}) and inefficient in superconductors, exhibiting
a strong systematic dependence of the vaporization energy $U_{v}$
on $T_{c}$. In contrast, melting of CDW order proceeds rapidly ($\tau_{m}=50\sim200$
fs) and more efficiently. A quantitative model describing the observed
systematic behavior in superconductors is proposed based on a phonon-mediated
quasiparticle (QP) bottleneck mechanism. In contrast, Fermi surface
disruption by hot QPs is proposed to be responsible for CDW state
melting.
\end{abstract}
\maketitle

Photoinduced phase transitions on the sub-picosecond timescale in
superconductors and other electronically ordered systems have attracted
increasing attention in recent years, partly because of the fundamental
desire to control collective states of matter with femtosecond laser
pulses, partly because of potential applications in ultrafast phase-change
memories, and partly because they may reveal some fundamental insight
into the mechanism of high-temperature superconductivity. However,
the details of how absorbed photons cause a change of state on ultrafast
timescales have so far not been investigated in detail. Recently,
a study on La$_{2-x}$Sr$_{x}$CuO$_{4}$ (LSCO) single crystals by
Kusar $et$ $al$.\cite{KusarPRL2008} showed that the energy required
to vaporise the superconducting (SC) condensate $U_{v}$ can be determined
with reasonable accuracy. These and other measurements on cuprate
superconductors since then\cite{Coslovich,Toda,NbN} give values of
$U_{v}$ which are sometimes significantly larger than the experimental
condensation energy $U{\mathrm{_{c}}}$, or the BCS theoretical values\cite{BCS},
opening the question of mechanism for the destruction of the condensate,
the dependence of $U_{v}$ on doping and on the critical temperature
$T_{c}$. It is also not clear if this occasional large discrepancy
between $U_{c}$ and $U_{v}$ is peculiar feature of cuprate superconductors,
is a consequence of a large gap in the density of states, or depends
on the detailed mechanism responsible for the formation of the low-temperature
ordered state.


To try and answer these questions, we present the first systematic
study of the dependence of $U_{\mathrm{v}}$ on doping and $T_{\mathrm{c}}$
for cuprates and compare our data with the iron pnictides (Ba(Fe$_{0.93}$Co$_{0.07}$)$_{2}$As$_{2}$
and SmFeAsO$_{0.8}$F$_{0.2}$), the conventional superconductor NbN
and two large-gapped charge-density wave systems K$_{0.3}$MoO$_{3}$
and TbTe$_{3}$. We made a special effort to determine the geometrical
factors and optical constants as accurately as possible in order to
achieve the best possible accuracy in determining $U{\mathrm{_{v}}}$.
We find that the observed systematic behavior reveals the mechanism
for breaking up the collective state in high-$T_{c}$ superconductors
to be relatively inefficient and phonon dominated. This is in stark
contrast to CDW systems, where the electronic destruction mechanism
is faster, more direct and efficient. We present a model to describe
the observed systematics in the cuprates and discuss the markedly
different mechanisms for the photonic destruction of the low-temperature
ordered state in the two types of system.

Our experiments were performed using a standard pump-probe technique
with a 800 nm laser with 50 fs pulses as described in ref.\cite{KusarPRL2008}.
A probe wavelength of 800 nm is chosen because it is a well-understood
probe wavelength for investigating the superconducting (SC) and pseudogap
(PG) response in pump-probe experiments and the best signal/noise
ratio as determined by previous studies. The 2-component response,
PG and SC, including lifetimes, $T$-dependence, anisotropy and doping
dependence have been previously studied at 800 nm and other pump/probe
wavelengths as well, including pump/probing in the THz region. The
results of these measurements are quantitatively self-consistent,
and have all been systematically discussed. The second reason for
using 800 nm (1.5 eV) is that the optical constants are known more
accurately at this wavelength better than any other, which improves
the accuracy of the determination of the vaporization energy. The
reason for not using far infrared (FIR) wavelengths in the gap region is
that even with transform-limited pulses, the pulses become too long
to be able to distinctly detect the SC component and distinguish it
from the PG. Hence THz or FIR pulses in the gap region are not suitable
for our purpose. Recently the electron distribution function thermalization
has been measured to be $\sim$ 60 fs in LaSCO and $\sim$ 100 fs
in YBCO\cite{Gadermaier}, while the quasiparticle (QP) signal used
for detection of the presence of the superconducting state in the
cuprates has a characteristic lifetime nearly two orders of magnitude
longer and can thus be easily distinguished from the PG and energy
relaxation which are both much faster. In order to accurately determine
the deposited energy, pump and probe laser beam diameters were accurately
measured with calibrated pinholes. The light penetration depth $\lambda_{\mathrm{op}}$
and reflectivity $R$, are obtained from published optical conductivity
and dielectric function data in each case \cite{CooperPRB1993}. The
measurements were performed on freshly cleaved YBa${_{2}}$Cu${_{3}}$O$_{7-\delta}$
single crystals with $T{_{\mathrm{c}}}$s of 90, 63 and 60 K, respectively
and Ca$_{x}$Y$_{1-x}$Ba$_{2}$Cu$_{3}$O$_{7-\delta}$ with $x=$
0.132 and $\delta=$ 0.072 and $T{\mathrm{_{c}}}$ $=$ 73 K. The
photoinduced reflectivity transient $\Delta$$R$/$R$ for YBa$_{2}$Cu$_{3}$O$_{6.5}$
well below $T_{c}$ (4 K) and above $T_{c}$ (at 68 K) are shown in
Fig. 1a), while the net superconducting signal, obtained by subtracting
the response at $T=$ 68 K from the response at $T=$ 4 K : $\Delta$$R$$\mathrm{_{sc}}$/$R$
= $\Delta$$R$/$R$|$_{T<T_{c}}$$-$$\Delta$$R$/$R$|$_{T>T_{c}}$
is plotted in Fig. 1b) for different laser pump pulse fluences. This
behavior is generic, so we have shown the raw data only for one sample.
$\Delta$$R$$\mathrm{_{sc}}$/$R$ shows a saturation plateau above
${\cal F}$ $\gtrsim$ 100 $\mu$J/cm$^{2}$\cite{KusarPRL2008},
which is a signature of the destruction of the SC state.

The normalized magnitude of $\Delta$$R$$\mathrm{_{sc}}$/$R$ for
different materials and different doping levels as a function of ${\cal F}$
is shown in Fig. 2. At low fluence $\Delta$$R$$\mathrm{_{sc}}$/$R$
is linearly dependent on ${\cal F}$ up to a certain threshold fluence
${\cal F}$$_{T}$, whereupon it either becomes saturated or continues
increasing, but with a different slope. To determine the vaporization
threshold fluence $\mathcal{F}_{T}$ from the data in Fig. 2, we use
the inhomogeneous excitation model \cite{KusarSupplement}. To take
into account carrier-density dependent change of $\Delta R/R$ arising
from $e-h$ asymmetry in the band structure, we have also included
a linear term in the fit to the data in Fig. 2, such that $\Delta R/\Delta R_{\mathrm{s}}=\Delta R_{\mathrm{sc}}/\Delta R_{\mathrm{s}}+\phi({\mathcal{F}})$.
Empirically, we find that in LSCO, $\phi\simeq0$ \cite{KusarPRL2008}
while in YBCO samples $\phi({\mathcal{F}})\simeq\phi_{0}\mathcal{F}$,
where $\phi_{0}$ is a constant. Note that this term has \textit{\emph{no
effect}} on the values of $\mathcal{F}_{T}$ obtained from the fits
to the data in Fig. 2. (As a check, $\mathcal{F_{T}}$ can also be
directly read out as the point where $\frac{\Delta R{}_{sc}(\mathcal{F})}{R}$
departs from linearity).

\begin{figure}[h]
 \includegraphics[width=8cm]{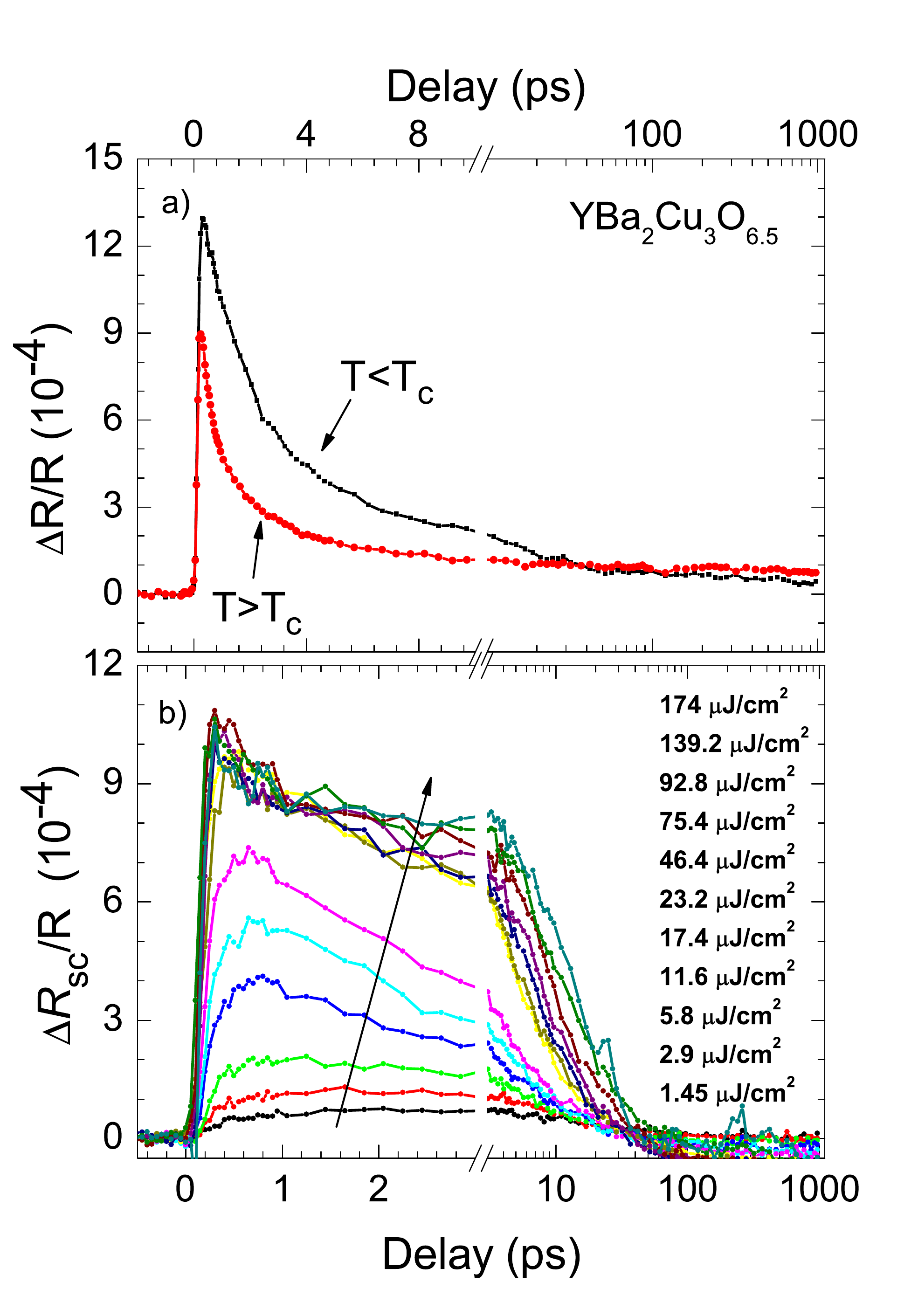} \caption{(Color online). a) $\Delta$$R$/$R$ for YBa$_{2}$Cu$_{3}$O$_{6.5}$
for $T$$<$$T_{c}$ (squares) and for $T$$>$$T_{c}$ (circles)
at excitation fluence ${\cal F}$ = 17.4 $\mu$J/cm$^{2}$. b) $\Delta$$R_{\mathrm{sc}}/R$
for YBa$_{2}$Cu$_{3}$O$_{6.5}$ as a function of time delay for
different $\mathcal{F}$, showing saturation above $46\mu$J/cm$^{2}$.
The arrow indicates the direction of increasing $\mathcal{F}$. The
condensate vaporization time $\tau_{v}$ is between 0.5 and 1 ps.
(At high fluences the PG signal interferes with the measurement, so
$\tau_{v}$ cannot be more accurately measured in YBaCuO).}
\end{figure}

\begin{figure}
\includegraphics[width=8cm]{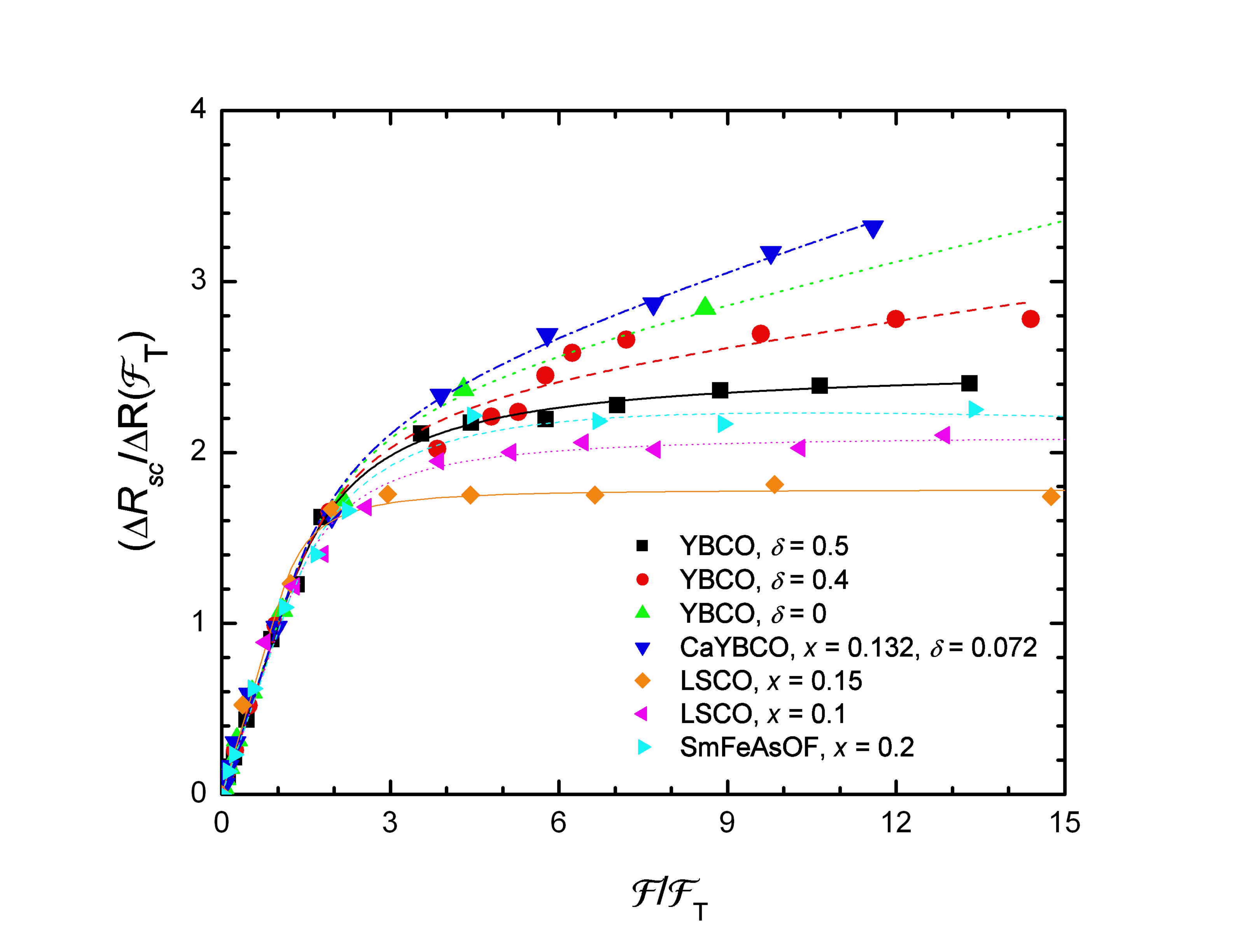} \caption{(Color online). $\Delta R_{\mathrm{sc}}$ as a function of normalised
fluence $\mathcal{F}/\mathcal{F}_{T}$. Note the linear response below
${\cal F}$$_{T}$ . }
\end{figure}
\begin{figure}[htb]
 \includegraphics[scale=0.3]{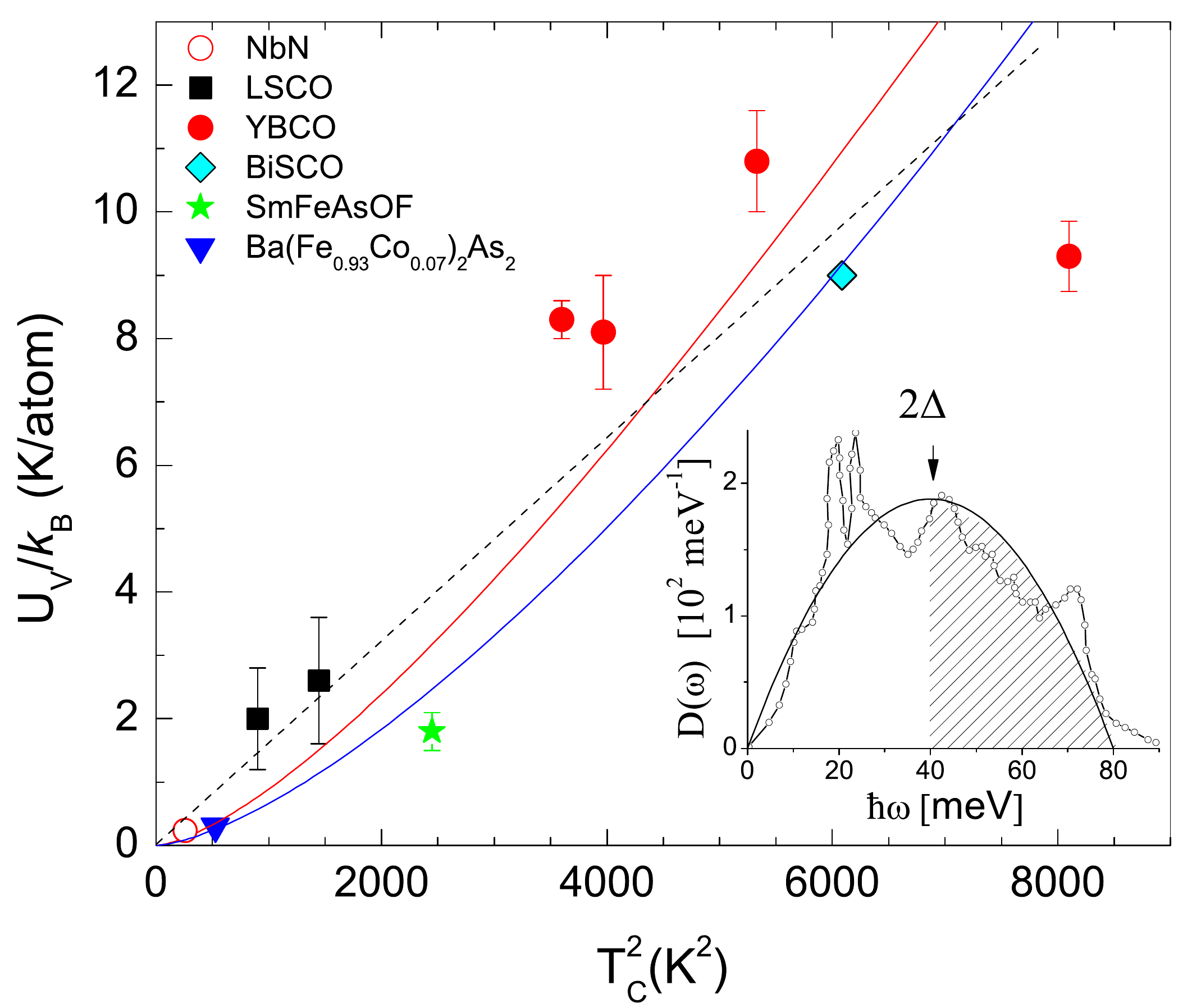} \caption{(Color online). $\Delta$$U$$\mathrm{_{v}}$ expressed in K per planar
Cu as a function of $T_{\mathrm{c}}$$^{2}$ for the cuprates. The
data for NbN, SmFeAsO$_{0.8}$F$_{0.2}$ and Ba(Fe$_{0.93}$,Co$_{0.07}$)$_{2}$As$_{2}$
are included for comparison (in K/Fe or K/Nb). The solid curve is
a plot of $U_{\mathrm{lost}}$ using $D(\omega_{\mathrm{ph}})$ appropriate
from Eq. (2) for YBaCuO (red) and the iron oxy-pnictide (blue). The
dashed line is a square law $U_{\mathrm{v}}=\eta T_{\mathrm{c}}^{2}$.
The insert shows the phonons with $\hbar\omega_{\mathrm{}}>2\Delta$
(shaded) which can break pairs. The measured phonon density of states
$D(\omega_{\mathrm{ph}})$ for YBCO\cite{PDOS} is approximated by
a parabola (line). }
\end{figure}

A plot of the vaporisation energy $U_{\mathrm{v}}=(1-R)\mathcal{F}_{\mathrm{T}}/\lambda_{\mathrm{op}}$
obtained for cuprates, the iron pnictides and NbN is shown in Fig.
3. We observe systematic increase of $U_{v}$ with $T_{c},$ giving
an approximate square power law dependence $U_{\mathrm{v}}=\eta T_{\mathrm{c}}^{2}$,
with $\eta\simeq1.5\times10^{-3}$K$^{-2}$atom$^{-1}$ irrespective
of whether the material is overdoped, optimally doped or underdoped.

In Table 1 we compare $U_{\mathrm{v}}$ with the condensation energy
$U_{\mathrm{c}}^{\mathrm{exp}}$ obtained from specific heat measurements.
In the cuprates, we notice a significant discrepancy between the magnitudes
of $U_{v}$ and $U{}_{c}^{exp}$. While $U_{\mathrm{c}}^{\mathrm{exp}}$
shows a distinct asymmetry between overdoped and underdoped materials
with the same $T_{c}$ \cite{LoramJPCS2001}, $U_{v}$ depends systematically
on $T_{c}$ irrespective of doping level. Thus $U_{v}$ and $U_{c}^{exp}$
show fundamentally different systematics.

\begingroup \squeezetable
\begin{table*}[t]
\begin{tabular}{cccccccccc}
\hline
Material  & $T_{\mathrm{c}}$ (K)  & ${\cal F}$$_{\mathrm{T}}$($\mu$J/cm$^{2}$)  & $\lambda_{\mathrm{op}}$(nm) & $U_{\mathrm{v}}$  & $U_{\mathrm{c}}^{\mathrm{exp}}$  &  & $U_{\mathrm{v}}$/$U_{\mathrm{c}}^{\mathrm{exp}}$  & $\tau_{v}$ (ps) & \tabularnewline
\hline
La$_{1.85}$Sr$_{\text{0.15}}$CuO$_{4}$ (opt) \cite{KusarPRL2008}  & 38.5  & 5.8$\pm$2.3 & 150 & 2.6$\pm$1.0 K/Cu  & 0.3 K/Cu \cite{KusarPRL2008}  &  & $>8.5$  & 0.9  & \tabularnewline
YBa$_{\text{2}}$Cu$_{3}$O$_{6.5}$ (ud)  & 60  & 13.1$\pm$0.5  & 90 & 8.3$\pm$0.3 K/Cu  & 0.38 K/Cu \cite{LoramJPCS2001}  &  & 22  & 0.5$\sim$ 1  & \tabularnewline
YBa$_{\text{2}}$Cu$_{3}$O$_{6.6}$ (ud)  & 63  & 12.1$\pm$1.3  & 85 & 8.1$\pm$0.9 K/Cu  & 0.45 K/Cu\cite{LoramJPCS2001}  &  & 18  &  & \tabularnewline
YBa$_{2}$Cu$_{3}$O$_{7}$ (opt)  & 90  & 10.8$\pm$0.6 & 66 & 9.3 $\pm$0.5 K/Cu  & 1.9 K/Cu \cite{LoramJPCS2001}  &  & 5  &  & \tabularnewline
Y$_{0.8}$Ca$_{0.2}$Ba$_{2}$Cu$_{3}$O$_{6.9}$ (od)  & 73  & 14.3$\pm$1  & 75 & 10.8$\pm$0.8 K/Cu  & 1.4K/Cu \cite{LoramJPCS2001}  &  & 7.7  &  & \tabularnewline
Bi$_{2}$Sr$_{2}$CaCu$_{2}$O$_{8+\delta}$ (ud)\cite{Toda} & 78 & 16$\pm$1 & 124 & 9$\pm0.8$ K/Cu & 1.8 &  & 5 &  & \tabularnewline
SmFeAsO$_{\text{0.8}}$F$_{\text{0.2}}$ \cite{MerteljPRL2009,MerteljCondoMat2010}  & 49.5  & 2.6$\pm$0.2  & 31 & 1.8$\pm$0.1 K/Fe  & -  &  & -  & 0.3-0.5  & \tabularnewline
Ba(Fe$_{0.93}$Co$_{0.07}$)$_{2}$As$_{2}$ & 23 & 0.43\textpm{}0.04 & 34 & 0.3$\pm$0.03 K/Fe & 0.15$\pm$0.02\cite{Hardy}  &  & 2 & 0.5  & \tabularnewline
NbN \cite{NbN}  & 16  & $U_{\mathrm{v}}=23$ mJ/cm$^{3}$  &  & 0.24 K/Nb  & 0.14K/Nb  &  & 1.7  &  & \tabularnewline
\hline
 & $T_{\mathrm{c}}$ (K)  & ${\cal F}$$_{\mathrm{T}}$($\mu$J/cm$^{2}$)  &  & $U_{m}$  & $U_{\mathrm{c}}^{\mathrm{exp}}$  & $U_{\mathrm{e}}^{\mathrm{theory}}$  & $U_{\mathrm{v}}/U_{\mathrm{e}}$  & $\tau_{m}$ & \tabularnewline
\hline
TbTe$_{3}$ \cite{Yusupov2010,TbTe3}  & 336  & 47$\pm$2  &  & 52$\pm$2 K/Tb  & -  & 40.6$\pm$1.7 K/Tb  & 1.3  & 0.2\cite{TbTe3} & \tabularnewline
K$_{\text{0.3}}$MoO$_{\text{3}}$ \cite{Kwok,Gruner,TomeljakPRL2009}  & 180  & 105$\pm$17.5  &  & 35$\pm$10K/Mo  & $>$10K/Mo  & 30$\pm$10 K/Mo\cite{TomeljakPRL2009}  & $1.1\sim3.5$  & 0.05-0.1\cite{TomeljakPRL2009}  & \tabularnewline
\hline
\end{tabular}\caption{$T_{\mathrm{c}}$, ${\cal F}$$_{\mathrm{T}}$, $U_{\mathrm{v}}$
and $U_{\mathrm{c}}$ for different materials. $U_{\mathrm{c}}$ and
$U_{\mathrm{v}}$ are calculated per metal atom of Cu (planar), Fe,
Nb, Tb or Mo. For LSCO and YBCO, 
the data for $U_{\mathrm{c}}$ are maximum experimentally measured
values\cite{LoramJPCS2001}. $\delta$ in YBCO was changed by annealing
in flowing oxygen, while $T_{\mathrm{c}}$s were determined from
magnetization measurements. For TbTe$_{3}$ and K$_{0.3}$MoO$_{3}$,
the electronic part of $U_{\mathrm{e}}$ of the condensation energy
is estimated using available values of $\Delta$ and $N_{0}$ \cite{TbTe3,Kwok}.
For YBa$_{2}$Cu$_{3}$O$_{7-\mbox{\ensuremath{\delta}}}$, the penetration
depths are\cite{CooperPRB1993}: 66$\pm$15 nm ($\delta=$ 0), 85$\pm$15
nm ($\delta=$ 0.4), 90$\pm$15 nm ($\delta=$ 0.5), and 75$\pm$15
nm for CaYBCO. }
\end{table*}

\endgroup

To understand this apparent discrepancy between the condensation and
vaporisation energy, let us now consider the mechanism of energy transfer
between photoexcited (PE) carriers and the condensate. After absorption
of a photon, the PE electrons and holes lose energy very rapidly through
{}``avalanche'' scattering with phonons on a timescale of $\tau_{E}<100$
fs for YBCO and $<60$ fs for LSCO \cite{Kabale}, creating a large
non-equilibrium phonon population in the process. The phonons whose
energy exceeds the gap $\hbar\omega_{\mathrm{ph}}>2\Delta$ can subsequently
excite QPs from the condensate, but these recombine again and so a
bottleneck occurs, in which high-frequency phonons are in temporarily
in quasi-equilibrium with QPs as described by Kusar $et$ $al$. for LaSrCuO\cite{KusarPRL2008}.
The process eventually causes the destruction of the condensate within
$0.5\sim1$ ps\cite{KusarPRL2008}.

The phonons with $\hbar\omega>2\Delta$ can contribute to the breakup
of the condensate, but the \textit{low frequency} phonons with $\hbar\omega<2\Delta$
also created in the initial avalanche do not have enough energy to
excite QPs, so do not contribute to the vaporisation process. This
lost energy to low-energy phonons is:
\begin{equation}
U_{\mathrm{lost}}=\int_{0}^{2\Delta}\Delta f(\omega)D(\omega)\hbar\omega d\omega
\end{equation}
 where $D(\omega)$ is the phonon density of states and $\Delta f(\omega)=f_{NE}(\omega)-f_{E}(\omega)$
is the difference between the non-equilibrium and equilibrium phonon
distribution functions. $\Delta f(\omega)$ is material-dependent
and depends on the electron phonon coupling, more precisely on the
Eliashberg coupling function $\alpha^{2}F(\omega)$, but is not a
very strong function of $\omega$\cite{Kabale}. To estimate the dependence
of $U_{lost}$ on $T_{c}$ from Eq. (1), we assume that $\Delta f(\omega)$
is constant, and approximate the experimental $D(\omega)$ by an inverted
parabola $D(\omega)=\frac{\alpha}{8}\omega(2\omega_{0}-\omega)$,
where $\alpha$ is a constant. Integrating Eq. (1), we obtain:
\begin{equation}
U_{\mathrm{lost}}=\alpha\Delta^{3}[\frac{2\hbar\omega_{0}}{3}-\frac{\Delta}{2}].
\end{equation}
Assuming a constant gap ratio $2\Delta/k_{\mathrm{B}}T_{\mathrm{c}}=R$,
with $R=4$, for YBCO, $\hbar\omega_{0}=40$ meV\cite{PDOS} extending
$D(\omega)$ to $80$ meV as shown in the insert to Fig. 3. Eq. (2)
gives the curve shown in Fig. 3. For iron pnictides, $\hbar\omega_{0}\simeq20$
\cite{PDOS1}, but the predicted variation of $U_{lost}$ on $T_{c}$
is not significantly different. We see that for the superconductor
series Eq. 2 predicts the dependence of $U_{v}$ on $T_{c}$ very
well, which is not surprising, considering the gross features of $D(\omega)$
do not vary significantly.

The total vaporisation energy is the sum of the condensation energy
and the lost energy, $U_{\mathrm{v}}=U_{\mathrm{c}}+U_{\mathrm{lost}}$.
Since for large gap systems $U_{\mathrm{v}}\gg U_{\mathrm{c}}^{\mathrm{exp}}$
we have $U_{v}\simeq U_{\mathrm{lost}}$, so $U_{c}^{exp}$ is thus
only a small contribution to $U_{v}$ explaining why the anomalous
doping dependence of $U_{c}^{exp}$ in the cuprates is not displayed
by $U_{v}$.

Comparing $U_{v}$ for NbN measured using the same technique (Table
I), $U_{\mathrm{v}}/U_{\mathrm{c}}^{\mathrm{exp}}=1.7$ is considerably
smaller than in the cuprates, which can now be understood in terms
of Eq. (1) and the insert to Fig. 3: When $2\Delta<<\hbar\omega_{\mathrm{Debye}}$
almost all phonons can excite QPs, so $U_{lost}\rightarrow0$, and
$U_{v}\simeq U_{c}$, in agreement with the data. For small-gap superconductors,
the optical method can be used to estimate the superconducting condensation
energy. The data on the two iron pnictides listed in Table I also
seem to follow the predicted behaviour in Fig. 3.

We now turn our attention to the data obtained using the same technique
on two different charge-density-wave materials TbTe$_{3}$ and K$_{\text{0.3}}$MoO$_{\text{3}}$.
Both have CDW gaps which are much larger than any phonon energy: $2\Delta_{CDW}\gg\hbar\omega_{D}$
($\Delta_{CDW}$=125 meV and 75 meV respectively\cite{Yusupov2008,Gruner}).
K$_{\text{0.3}}$MoO$_{\text{3}}$ is considered a prototypical Peierls
system with a full gap in the electronic density of states opening
at the metal-insulator transition\cite{Gruner} Moreover, in KMo$_{0.3}$O$_{3}$
the electron-phonon coupling constant ($\lambda_{e-p}\simeq0.35$)
is comparable with the cuprates ($\lambda_{e-p}\simeq0.25$ for YBaCuO
and $\lambda_{e-p}=0.5$ for LaSCO respectively)\cite{Gadermaier,Gruner},
so with a phonon-mediated CDW melting mechanism we would expect a
very large $U_{m}/U_{c}$ ratio.  Instead, it is significantly \emph{smaller}
than in the large-gapped cuprates, with $U_{m}/U_{c}^{exp}\simeq$
3.5. In TbTe$_{3}$ the ordered state opens a partial gap only along
certain directions on the Fermi surface, the rest remaining ungapped\cite{TbTe3}.
Thus, if the same phonon-mediated vaporisation mechanism were operative,
the ungapped CDW system should behave very differently to KMo$_{0.3}$O$_{3}$
on one hand, and the superconductors on the other. Comparing $U_{\mathrm{v}}$
with the electronic contribution to the CDW condensation energy\cite{Gruner,TomeljakPRL2009}:
$U_{\mathrm{e}}=N_{0}\Delta^{2}(\frac{1}{2}+log\frac{2E_{\mathrm{F}}}{\Delta})$,
where $E_{\mathrm{F}}$ is the Fermi energy \cite{TbTe3}, we see
in Table I that the predicted ratios are $U_{\mathrm{m}}/U_{\mathrm{e}}\simeq1.3$
for TbTe$_{3}$ and $1.1$ for K$_{0.3}$MoO$_{3}$, confirming that
$U_{lost}\rightarrow0$ and implying that phonons are not significantly
involved in the CDW gap destruction and CDW melting process. Thus,
in spite of similar \emph{e-p} interaction strength and very large
gaps, $U_{m}/U_{c}$ is consistently smaller in the CDWs than $U_{v}/U_{c}$
in cuprates, highlighting the very different mechanism for the destruction
of the low-temperature CDW state.

The reason is fundametal: Unlike in superconductors, PE carriers in
CDW systems can cause a direct rapid disturbance of the Fermi surface
\cite{Yusupov2010} on a timescale shorter than the QP recombination
time. It is known experimentally\cite{DopedKMO} and theoretically\cite{Maki}
that the CDW state is extremely sensitive to charge imbalance, which
disrupts the Fermi surface (FS) nesting. In photoexcitation experiments,
overall charge neutrality is preserved, but immediately after photon
absorption, even the slightest electron-hole band asymmetry causes
a transient shift of the chemical potential, causing a disturbance
$\lyxmathsym{\textgreek{d}}q$ of the FS. This causes a destruction
of the nesting at $k_{F}$ and a suppression of the divergence of
the electronic susceptibility at $2k_{\mathrm{F}}$ causing the electronic
charge density wave to rapidly melt\cite{Maki,Schmitt}. A fast destruction
time in CDWs is experimentally confirmed: in K$_{0.3}$MoO$_{3}$,
$\tau_{m}=50\sim100$ fs\cite{TomeljakPRL2009}, which is an order
of magnitude less than $\tau_{v}\sim0.9$ ps in LaSCO, which has a
similar \emph{e-p} interaction strength. Indeed, quite generally,
the melying time $\tau_{m}$ is significantly shorter than the $\tau_{v}$
in superconductors, as shown in Table I (not all risetimes are reliably
measured so far, so only the unambiguous ones are listed). With $\tau_{m}=50\sim100$
fs, the ions can just barely move into new equilibrium positions within
this time, (the $1/4$-period of the amplitude mode in K$_{\text{0.3}}$MoO$_{\text{3}}$
is $\sim$ 150 fs).\cite{TRED}

We conclude that photon-induced change of state in superconductors
and CDW systems relies on very different mechanisms, occuring on different
timescales and requiring different amounts of energy. Systematic and
experimentally precise measurements of the dependence of $U_{v}$
on doping and on $T_{c}$ in cuprates, pnictides, low-$T_{c}$ superconductors
show that \foreignlanguage{british}{vaporisation} can be described
by the established Rothwarf-Taylor QP recombination mechanism, in
a process where energy transfer to the condensate is slowed down and
made inefficient by the phonon-QP relaxation bottleneck. In CDW systems
on the other hand, the dominant mechanism for the destruction of the
ordered state is electronic, so in spite of the fact that the CDW
gaps $2\Delta_{CDW}$ are typically much larger than the highest phonon
frequencies, and electron-phonon coupling strengths are similar as
in cuprates, the destruction is faster and more direct.

\bibliographystyle{unsrt} 

\end{document}